\documentclass[9pt,twocolumn,twoside]{osajnl}

\journal{ol} 

\setboolean{shortarticle}{true}

\title{Fast phase cycling in non-collinear optical two-dimensional coherent spectroscopy}

\author[1]{Maria F. Munoz}
\author[1]{Adam Medina}
\author[2]{Travis M. Autry}
\author[3]{Galan Moody}
\author[4]{Mark E. Siemens}
\author[5]{Alan D. Bristow}
\author[6]{Steven T. Cundiff}
\author[1,*]{Hebin Li}
\affil[1]{Department of Physics, Florida International University, Miami, Florida 33199, USA}
\affil[2]{National Institute of Standards and Technology, Boulder, Colorado 80305, USA}
\affil[3]{Department of Electrical and Computer Engineering, University of California Santa Barbara, Santa Barbara, California 93106, USA}
\affil[4]{Department of Physics and Astronomy, University of Denver, Denver, Colorado 80210, USA}
\affil[5]{Department of Physics and Astronomy, West Virginia University, Morgantown, West Virginia 26506-6315, USA}
\affil[6]{Department of Physics, University of Michigan, Ann Arbor, Michigan 48109, USA}

\affil[*]{Corresponding author: hebin.li@fiu.edu}

\doi{}

\begin{abstract}
As optical two-dimensional coherent spectroscopy (2DCS) is extended to a broader range of applications, it is critical to improve the detection sensitivity of optical 2DCS. We developed a fast phase-cycling scheme in a non-collinear optical 2DCS implementation by using liquid crystal phase retarders to modulate the phases of two excitation pulses. The background in the signal can be eliminated by combining either two or four interferograms measured with a proper phase configuration. The effectiveness of this method was validated in optical 2DCS measurements of an atomic vapor. This fast phase-cycling scheme will enable optical 2DCS in novel emerging applications that require enhanced detection sensitivity. 
\end{abstract}
\setboolean{displaycopyright}{false}

\begin{document}
\maketitle
Optical two-dimensional coherent spectroscopy (2DCS) has found important applications in systems such as atomic vapors \cite{Tian2003,Tekavec2007,Dai2010,Dai2012,Li2013,Gao:16,PhysRevLett.120.233401,Yu2019,Yu2018,Bruder2019}, semiconductor quantum wells \cite{Li2006a,Stone2009a,Turner2012,Singh2013,PhysRevLett.112.097401,Nardin2014} and dots \cite{PhysRevB.87.041304,Moody2013,Moody2013a,Moody2013b}, 2D materials \cite{Moody2015,Titze2018,martin2018encapsulation}, perovskites \cite{Monahan2017,Richter2017,Jha2018,Thouin2018,Nishida2018,Titze2019}, and photosynthesis \cite{Brixner2005,Engel2007,Collini2010}. In optical 2DCS, only nonlinear signal corresponding to specific excitation quantum pathways is desired to be measured; nonlinear signals from other excitation pathways, fluorescence, and non-resonance scatter of excitation laser beams are considered as background, diminishing the signal-to-noise ratio (SNR) and ultimately the sensitivity of the technique. These backgrounds might not be removable by methods such as phase matching, polarization control, fitting, etc. New emerging applications of optical 2DCS may pose challenges in the detection sensitivity. For instance, measurements on atomically thin two-dimensional materials are complicated by strong laser scatter from the substrate in comparison to a weak nonlinear signal from the sample itself. Other examples include cold atoms, single molecules, biological samples, etc., involving strong background or weak signal, or both. To extend optical 2DCS to a broader range of applications, it is essential to have an effective method to reduce the background and increase SNR and detection sensitivity. 

Optical 2DCS has been implemented in both collinear and non-collinear approaches. In the collinear geometry where the excitation pulses copropagate, the signals from all excitation pathways emit in the same direction. A phase cycling procedure can be performed to selectively detect the desired signal corresponding to a specific pathway \cite{Shim2007,Tekavec2007,Wagner2005}. The phase of each excitation pulse is toggled to change the phase of the signal from certain pathways but not others. The desired signals from specific pathways can be isolated by coherently subtracting and/or adding a proper combination of spectra acquired with different phase configurations of excitation pulses. Alternatively, the excitation pulses can be frequency tagged by acousto-optic modulators (AOMs) and a specific signal can be detected by a lock-in amplifier at a proper mixing of AOM frequencies \cite{Nardin2013,Tekavec2007}. This frequency selection scheme can be understood as dynamic pulse-to-pulse phase cycling \cite{Nardin2013}. Specific nonlinear signals can also be isolated in frequency-comb based 2DCS approaches \cite{PhysRevLett.120.233401,Lomsadze2017b,Lomsadze2018} with enhanced frequency resolution. 

In the non-collinear geometry, three parallel excitation beams are aligned to the three corners of a square and converge on the sample by a lens. To ensure phase stability between the pulses, they propagate through common optics so they are subject to the same path fluctuations \cite{Brixner2004,Cowan2004,Turner2011}, or their relative phases are actively stabilized by using interferometric error signals \cite{Bristow2009,Zhang2005}. One implementation of active stabilization is known as ``multidimensional optical nonlinear spectrometer'' (MONSTR) \cite{Bristow2009}, where four pulses are derived from three nested Michelson interferometers. The four pulses are fully phase-locked and their relative delays can be scanned up to 1 ns, allowing multidimensional coherent spectroscopy on systems with long coherence and lifetime. The signal from specific pathways emits in a particular phase-matching direction, and thus can be spatially filtered to reject most background signals from other pathways. However, the isotropic fluorescence signal and the scatter of laser beams can still present in the signal emission direction. To remove the background, one method is to block and unblock different beams sequentially by using either shutters \cite{Brixner2004b} or choppers \cite{Augulis2011} in order to measure different background contributions. A phase-cycling scheme in the non-collinear geometry is to vary the phase by stepping delay stages \cite{Bristow2009}. These approaches are relatively slow and might not remove background contributions from all beams. 

In this letter, we report a fast phase cycling method in non-collinear optical 2DCS. Based on the original MONSTR implementation, contrast to stepping delay stages, two liquid crystal phase retarders are inserted in the paths of two beams to modulate the phases of two excitation pulses. By adding and subtracting a series of interferograms measured with a proper phase configuration, we can cancel out the background while keeping the signal in the resulting spectrum. A four-step phase cycling procedure can eliminate most background scatter and fluorescence due to all excitation pulses. We demonstrate the effectiveness of the phase-cycling technique with 2DCS measurements in a potassium (K) atomic vapor, enhancing the SNR by a factor of 5.   

\begin{figure}[htbp]
\centering
\includegraphics[width=\linewidth]{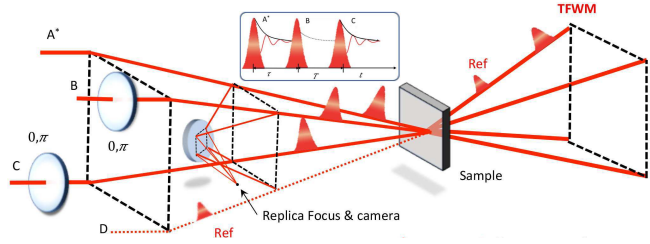}
\caption{Schematic of the non-collinear optical 2DCS setup with two liquid crystal phase retarders inserted in the beam paths. The inset shows the pulse sequence for rephasing 2DCS.}
\label{fig:setup}
\end{figure}

The schematic of the 2DCS experiment is shown in Fig. \ref{fig:setup}. Four phase-locked pulses are aligned in the box geometry and are focused onto the sample by a lens. Three of them, labeled $A^*$, $B$, and $C$, generate a third-order transient four-wave mixing (TFWM) signal in the propagation direction of the fourth pulse. The time delays are denoted as $\tau$ between the first two pulses, $T$ between the second and third, and $t$ the emission time relative to the third pulse. The TFWM signal is detected through spectral interferometry by using the fourth pulse as a reference (local oscillator). The reference pulse can either bypass or go through the sample. The latter is used in this experiment and the reference pulse is at least 1000 times weaker in intensity than the excitation pulses so the excitation by the reference pulse is negligible. The signal is recorded as a function of time delays and Fourier transformed into the frequency domain to generate a 2D or 3D spectrum. Various types of 2D and 3D spectra can be obtained depending on the pulse time ordering and which delays are scanned \cite{Bristow2009}. Here we consider only the rephasing 2D spectrum obtained with the pulse sequence shown in Fig. \ref{fig:setup} but the technique applies to other 2D and 3D spectra as well. 

\begin{figure}[htbp]
\centering
\includegraphics[width=\linewidth]{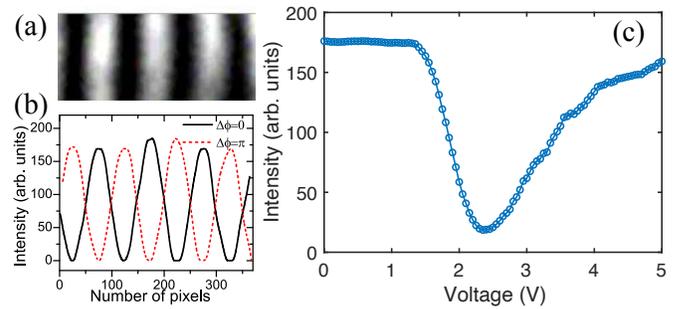}
\caption{(a) Interference fringe pattern of pulses $A$ and $B$. (b) Horizontal slices corresponding to zero (black) and $\pi$ (red dash) phase shift. (c) The intensity at a pixel as a function of the applied voltage on the liquid crystal retarder. }
\label{fig:calibration}
\end{figure}

To control the phase of pulses, two liquid crystal phase retarders are inserted in the paths of pulses $B$ and $C$ (any two of the three excitation pulses will work) \cite{Wahlstrand:19}. The laser pulses are linearly polarized and the polarization direction is aligned to the slow axis of the liquid crystals so that they can modulate the phase without affecting the polarization or intensity.We perform experiments with linear polarized light, but circular polarization could be obtained by placing quarter waveplates after the phase retarders. The phase of each pulse can be changed by varying the amplitude of a 2-kHz square wave voltage to the liquid crystals. This phase change can be measured from the interference fringes between two pulses when they converge in both space and time at the focal point. For convenience, the fringes are measured at a replica of the focal point that is formed by placing a beam splitter between the lens and the sample. The replica is imaged by an objective lens and a CCD camera. As an example, a region of the interference fringes between pulses $A$ and $B$ are shown in Fig. \ref{fig:calibration}(a). A horizontal slice of the pattern is plotted as the black curve in Fig. \ref{fig:calibration}(b). As the phase of pulse $B$ varies, the interference fringes move along the horizontal direction. The bright and dark stripes swap their positions when the phase is changed by $\pi$, as indicated by the red dash line in Fig. \ref{fig:calibration}(b). To determine the voltage that is required to induce a $\pi$ phase shift, we find the central position of a bright stripe in Fig. \ref{fig:calibration}(a) and measure the intensity at the same pixel as the voltage is varied from 0 to 5 V. A typical measurement is shown in Fig. \ref{fig:calibration}(c) where the intensity is plotted as a function of the voltage. The intensity is maximum at zero voltage and minimum when the phase shift is $\pi$. The voltage corresponding to the minimum intensity will be used to apply a $\pi$ phase shift (relative to the phase at zero voltage) to a pulse in the phase cycling procedure. Both liquid crystal phase retarders are calibrated in this way before 2DCS measurements.  

The third-order TFWM nonlinear signal measured in our experiment can be written as 
\begin{equation}
    E^{(3)}_{TFWM}(\omega) \propto \chi^{(3)} \hat{E}^*_A(\omega) \hat{E}_B(\omega) \hat{E}_C(\omega), 
\end{equation}
where $\chi^{(3)}$ is the third-order nonlinear susceptibility and $\hat{E}_{A,B,C}(\omega)=\tilde{E}_{A,B,C}(\omega) e^{-i\omega t^*}e^{i\phi_{A,B,C}}$ are the electric fields of three pulses. Here $\tilde{E}_{A,B,C}(\omega)$ are the pulse envelopes, $\phi_{A,B,C}$ are the phases of each pulse, and $t^*$ is time. Therefore, the TFWM signal becomes 
\begin{equation}
    E^{(3)}_{TFWM}(\omega) \propto \chi^{(3)}\tilde{E}^*_{A}\tilde{E}_{B}\tilde{E}_{C}e^{-i\omega t}e^{-i\phi_{A}}e^{i\phi_{B}}e^{i\phi_{C}}.
\end{equation}
The TFWM signal depends on the phases of all three excitation pulses as $\phi_{TFWM}=-\phi_A+\phi_B+\phi_C$. If any of the excitation pulses, for example, pulse $B$, has a $\pi$ phase shift, both TFWM signal  $E^{(3)}_{TFWM}$ and pulse $\hat{E}_B$ (thus its scatter) have a sign change ($e^{i\pi}=-1$) while the fields of other pulses and their scatter remain the same. If two pulses each have a $\pi$ phase shift, the TFWM signal does not change sign. By toggling the phase shift between 0 and $\pi$ of two excitation pulses, a phase cycling procedure can be implemented to eliminate the background.

\begin{table}[htbp]
\centering
\caption{\bf Phase cycling operations by toggling the phases of two pulses.}
\begin{tabular}{c||c|cccc}
\hline
Operation & $E_{tot}$ & $E^{(3)}_{TFWM}$ & $E_{S,A}$ & $E_{S,B}$ & $E_{S,C}$ \\
\hline
$\Delta\phi_B=0,\  \Delta\phi_C=0$ & $S_1$ & $+$ & $+$ & $+$ & $+$ \\
$\Delta\phi_B=\pi,\  \Delta\phi_C=0$& $S_2$ & $-$ & $+$ & $-$ & $+$ \\
$\Delta\phi_B=\pi,\  \Delta\phi_C=\pi$ & $S_3$ & $+$ & $+$ & $-$ & $-$ \\
$\Delta\phi_B=0,\  \Delta\phi_C=\pi$ & $S_4$ & $-$ & $+$ & $+$ & $-$ \\
\hline
\end{tabular}
  \label{tab:operations}
\end{table}

\begin{figure*}[htb]
\centering
\includegraphics[width=\linewidth]{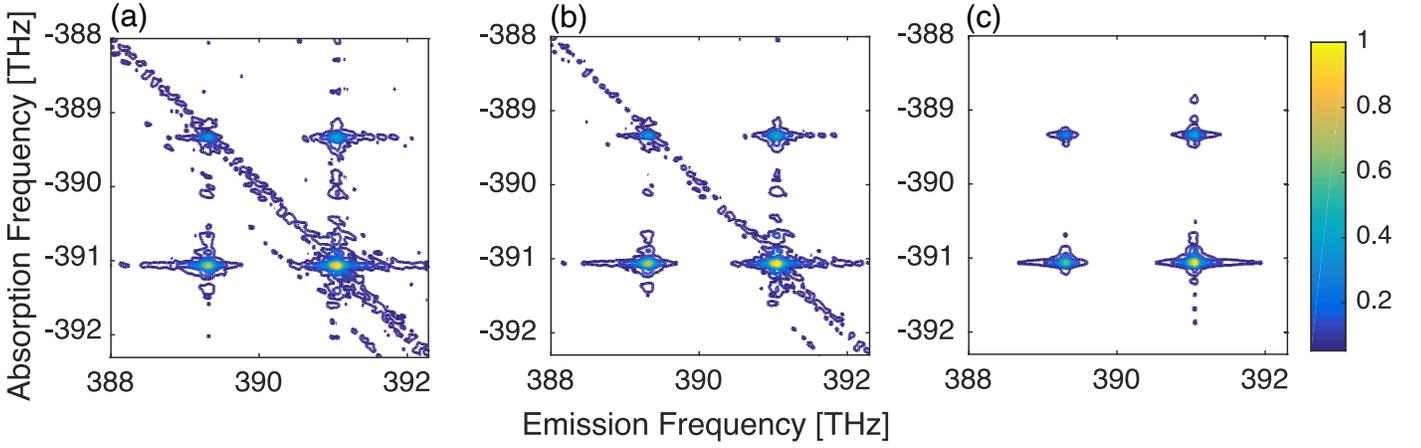}
\caption{2D rephasing spectra of K vapor measured with (a) no phase cycling, (b) two-step phase cycling, and (c) four-step phase cycling. The amplitude is plotted and the maximum is normalized to one in each spectrum.}
\label{fig:2Dspectra}
\end{figure*}

Considering the background, the total signal $E_{tot}$ measured in the experiment includes the TFWM signal and the scatter of each pulse, that is
\begin{equation}
    E_{tot}=E^{(3)}_{TFWM}+E_{S,A}+E_{S,B}+E_{S,C}. 
\end{equation}
To eliminate the background and extract the TFWM signal, we perform a phase cycling procedure with four operations, as shown in Table \ref{tab:operations}. In the first step zero voltage is applied to the liquid crystal phase retarders. As a reference, the phase shifts in pulses $B$ and $C$ are considered zero ($\Delta\phi_B=0,\  \Delta\phi_C=0$ ) and the signs of signal fields are $+$. The measured total signal is denoted as $S_1$. In the second step, a calibrated voltage is applied to the retarder for pulse $B$ to induce a phase shift $\Delta\phi_B=\pi$. The signs of $E^{(3)}_{TFWM}$ and $E_{S,B}$ flip and the resulting total signal is $S_2$. Subtracting the signals measured in these two operations gives $S_1-S_2=2(E^{(3)}_{TFWM}+E_{S,B})$. That is, the background due to pulses $A$ and $C$ can be eliminated by using only one liquid crystal to change the phase once. This two-step phase cycling might be sufficient in some cases where pulses $A$ and $C$ are the main sources of background. To further eliminate the background due to pulse $B$, the second liquid crystal phase retarder and two more operations are needed. In step 3, both liquid crystal phase retarders are applied voltage to have phase shifts $\Delta\phi_B=\pi,\  \Delta\phi_C=\pi$. The sign of $E^{(3)}_{TFWM}$ changes back to $+$ while $E_{S,B}$ and $E_{S,C}$ have a $-$ sign. The signal measured in step 3 is denoted as $S_3$. In the last step, the liquid crystals set the phase shifts to $\Delta\phi_B=0,\  \Delta\phi_C=\pi$ and the operation flips the signs for $E^{(3)}_{TFWM}$ and $E_{S,B}$ compared to that in the previous step. The resulting signal is $S_4$. Using the signals obtained in these four operations, we can calculate $S_1-S_2+S_3-S_4=4E^{(3)}_{TFWM}$. Therefore, the TFWM signal can be extracted without the background of all excitation pulses after the four-step phase cycling as 
\begin{equation}
    E^{(3)}_{TFWM}=\frac{1}{4}(S_1-S_2+S_3-S_4). 
\end{equation}

To test its effectiveness, this phase-cycling technique is applied to  optical 2DCS on atomic K vapor. The K vapor is contained in a thin vapor cell made of a titanium body and two sapphire windows forming a gap of about 20 $\mu$m. The cell is also filled with an argon buffer gas and heated to 165 $^\circ$C during the experiment to broaden the linewidth. The excitation laser produces 35-fs pulses at a 5-KHz repetition rate with a central wavelength of 775 nm. The laser bandwidth covers both $D_{1}$ (389.29 THz, $4^2 S_{1/2}  \leftrightarrow 4^2 P_{1/2} $) and $D_{2}$ (391.02 THz, $4^2 S_{1/2}  \leftrightarrow 4^2 P_{3/2} $) transitions. The K vapor can be considered as a three-level $V$ energy level system and its 2D spectra have been reported before \cite{Dai2010}. For the purpose of demonstration, we intentionally introduce excess noise in the experiment by measuring at a sample spot that has a stronger scatter. 

Measured rephasing 2D spectra of K vapor are shown in Fig. \ref{fig:2Dspectra}. The spectrum features two diagonal peaks corresponding to the $D_{1}$ and $D_{2}$ transitions and two off-diagonal peaks due to the coupling of the two transitions \cite{Dai2010}. The 2D spectrum obtained without phase cycling, as shown in Fig. \ref{fig:2Dspectra}(a), has significant background distributed primarily along the diagonal direction. This is expected since the absorption and emission frequencies are identical for the scatter as the main background. The two-step phase-cycling procedure is performed and $(S_1-S_2)/2$ is used to generate a 2D spectrum, and the resulting spectrum is shown in Fig. \ref{fig:2Dspectra}(b). The background is reduced but not completely eliminated since the two-step phase cycling removes only the scatter of pulses $A$ and $C$ but not pulse $B$. To eliminate the scatter of all pulses, the four-step phase cycling procedure is performed to obtain $(S_1-S_2+S_3-S_4)/4$ for generating 2D spectra. The obtained 2D spectrum, as shown in Fig. \ref{fig:2Dspectra}(c), has no background visible along the diagonal line. The SNR is enhanced by a factor of 5 compared to the spectrum obtained without phase cycling. This experiment confirms the effectiveness of the phase-cycling method and shows that the four-step phase cycling eliminates scatter of all pulses while the two-step procedure only partially removes the background. 

A phase-cycling scheme was previously implemented \cite{Bristow2009}. The phase modulation of a pulse was achieved by varying the phase using a mechanical delay line. In the MONSTR, the time delay can only increment a distance commensurate with an integer number of Helium-Neon (HeNe) reference laser (632.8 nm) fringes to ensure the phase stabilization. Since the fs excitation pulses and the HeNe reference laser have different wavelengths, the pulse delay needs to be varied more than a half cycle to achieve a phase shift close to $\pi$. For example, when the fs excitation laser is tuned to 768 nm for measurements on K atoms, the shortest increment of the delay stage is 6 HeNe fringes corresponding to approximately $5\pi$ at the fs excitation laser wavelength. This approach affects the precision of the pulse time delay and may not provide an exact $\pi$ phase shift, especially when the fs laser wavelength is not ``convenient''. Moreover, the data acquisition time is significantly longer since moving the delay stage is the most time-consuming operation in the 2DCS scan. In comparison, the liquid crystal phase retarder can be calibrated to shift the phase exactly by $\pi$. The phase retarder has a fast switching speed with a typical rise and fall time of 34 ms and 360 $\mu$s, respectively. The extra dispersion due to the liquid crystal phase retarder was not a factor in these experiments, but it could be a concern for ultrashort sub-10-fs pulses. In this case, the dispersion can be pre-compensated by negatively chirping the input excitation pulses.  

In summary, we implemented a fast phase-cycling scheme in non-collinear optical 2DCS. The phase of two excitation pulses can be modulated by two liquid crystal phase retarders. By toggling the phase between $0$ and $\pi$, a series of interferograms with a proper phase configuration were obtained to eliminate the background while keeping the signal. The method can effectively eliminate the background and improve the SNR in 2D spectra, as demonstrated by 2DCS measurements in a K atomic vapor. This fast phase-cycling technique will broaden the scope of optical 2DCS applications, enabling 2DCS measurements in samples that have strong background or weak signals, or both.




\medskip

\noindent\textbf{Funding.} National Science Foundation (PHY-1707364).

\medskip

\noindent\textbf{Disclosures.} The authors declare no conflicts of interest.

\bibliography{phasecycling}


\end{document}